\documentclass{optica-article}

\journal{opticajournal} 

\articletype{Research Article}

\usepackage{lineno}
\usepackage{siunitx}

\begin{document}

\title{Focusing through coupled multimode fibers for miniaturized single-atom tweezer arrays}

\author{Benjamin P. Maddox\authormark{1,*,\dag} and Daniel K. Ruttley\authormark{1,*,\ddag}}

\address{\authormark{1}Department of Physics, Durham University, South Road, Durham, DH1 3LE, United Kingdom\\
\authormark{*}The authors contributed equally to this work.\\
\email{\authormark{\dag}benjamin.p.maddox@durham.ac.uk}
\email{\authormark{\ddag}daniel.k.ruttley@durham.ac.uk}}

\begin{abstract*} 
Real-world deployment of quantum technologies requires systems in low SWaP-C packages.
Quantum platforms based on ultracold atoms typically trap and control atoms with optical tweezers using light focused through bulky objective lenses, which are costly and sensitive to misalignment.
Multimode fibers (MMFs) offer a compact route to focusing light, but their use in forming optical tweezers for cold atoms has been limited to a single fiber.
Here we use a $2 \times 1$ MMF fused coupler to obtain two independent control paths sharing a common distal facet.
We form near-diffraction-limited tweezers through each channel, suppress interference by applying a small frequency offset between the channels, and demonstrate that the tweezer separation is insensitive to misalignment of the input coupling. 
This work provides a route to reconfigurable single-atom arrays in low SWaP-C environments.
\end{abstract*}

\section{Introduction}
Quantum technologies will enable new capabilities in sensing, communication, and information processing~\cite{Hanzo2025}.
Realizing these applications outside laboratory environments will require quantum systems to be integrated into low size, weight, power, and cost (SWaP-C) packages. Quantum networks~\cite{Kimble2008}, for example, enable applications including device-independent quantum key distribution~\cite{Acin2007,Nadlinger2022,Zhang2022b,Arslan2026} and distributed quantum computing~\cite{Cuomo2020}. However, long-distance links will require chains of quantum repeaters~\cite{Briegel1998} to be distributed along communication lines.
Repeater architectures based on individually trapped ions~\cite{Stephenson2020,Krutyanskiy2023,OReilly2024,Liu2026a} and neutral atoms~\cite{Welte2021,VanLeent2022} are particularly promising for these applications, where matter qubits act as quantum memories. 
Reducing the size of such systems, however, places stringent constraints on the optical components required to trap, control, and readout the constituent quantum particles.
Current implementations require objective lenses which are bulky, costly, and sensitive to misalignment, which limits their deployment in real-world environments.

Optical fibers offer an attractive route towards compact optical systems because of their small physical size, flexibility, and compatibility with constrained and cryogenic environments. Several demonstrations have explored fiber-based trapping and control of individual atoms. Garcia~\textit{et al.}~\cite{Garcia2013} demonstrated trapping and readout of single rubidium atoms using an optical tweezer formed by an aspheric lens attached to the end of a single-mode fiber (SMF). Similar functionality was subsequently demonstrated using 3D-printed lenses fabricated directly on the end of SMFs, with the printed structures compatible with vacuum operation~\cite{Ruchka2022}. More recently, Safari~\emph{et al.}~\cite{Safari2026} demonstrated enhanced atom-light coupling in a compact system by focusing light from an SMF with an in-vacuum parabolic mirror. While these approaches provide high optical access in constrained environments, the fixed spatial mode of an SMF limits the dynamic control of the trapping potential. In particular, independent control of the amplitude and position of optical tweezers, enabling fast reconfiguration~\cite{Miroshnychenko2006,Lee2016,Barredo2016,Endres2016} and switching~\cite{Labuhn2016}, are required to implement the local entanglement operations necessary for a quantum repeater using Rydberg gates.

Multimode fibers (MMFs) provide a route to overcome this limitation by allowing the spatial structure of the output field to be controlled through the input field~\cite{Cao2023}. 
MMFs support a large number of spatial modes while retaining the small footprint and flexibility of SMFs, making them attractive for applications requiring compact and reconfigurable optical systems.
This has motivated extensive work on controlling light through MMFs~\cite{Flusberg2005,Cizmar2011,Bianchi2012,Gu2014,Gu2015,Gigan2022,Amitonova2024}, with applications including medical endoscopy~\cite{Cao2023,Amitonova2024}, microparticle manipulation~\cite{Bianchi2012}, and trapping and imaging ultracold atoms~\cite{Vitrant2021,Garcia2022}.

Previous demonstrations have predominantly used single-input, single-output MMF configurations, in which multiple optical fields must be generated through the same control channel. For example, Garcia~\textit{et al.}~\cite{Garcia2022} demonstrated trapping and imaging of atoms through an MMF, but generated multiple tweezers from a common optical channel. Such an arrangement can introduce interference between tweezers and prevents fast, independent control of their position and amplitude \cite{caravaca2017single}. 
Many endoscopy schemes implement rasterisation, by time multiplexing individual spots, but this approach is challenging for atomic control because typical refresh rates of devices used to shape light are \si{\kilo\hertz}-scale, comparable to typical trap frequencies of trapped atoms.
Multimode-fiber fused couplers offer an alternative architecture in which multiple input channels share a common output, providing independent control of several optical fields at the same compact distal interface.

Here, we demonstrate the use of a multimode-fiber fused coupler to generate independently controllable optical tweezers for compact quantum systems. We use a $2 \times 1$ MMF coupler in a two-input, one-output configuration and independently shape the spatial modes launched into each input. We first demonstrate control of the output field through a single channel and then use both channels to generate two optical tweezers at independently chosen positions. We characterize the resulting tweezers and show that they exhibit minimal crosstalk while maintaining their quality over a range of separations. Crucially, the two input channels retain independent frequency  and amplitude control, allowing their relative detuning to be chosen arbitrarily and thereby suppressing interference on the timescale relevant to atomic dynamics. These results establish MMF couplers as a compact platform for generating independently addressable optical fields and provide a route towards fiber-based trapping and control of ultracold atomic arrays.

\section{Experimental setup} \label{sec:setup}
In this section, we introduce the optical setup used in this work and review how the transmission of light through a MMF can be characterized and exploited in order to produce optical tweezers.

\subsection{Optical system}
\begin{figure}[]
\centering\includegraphics[width=\textwidth]{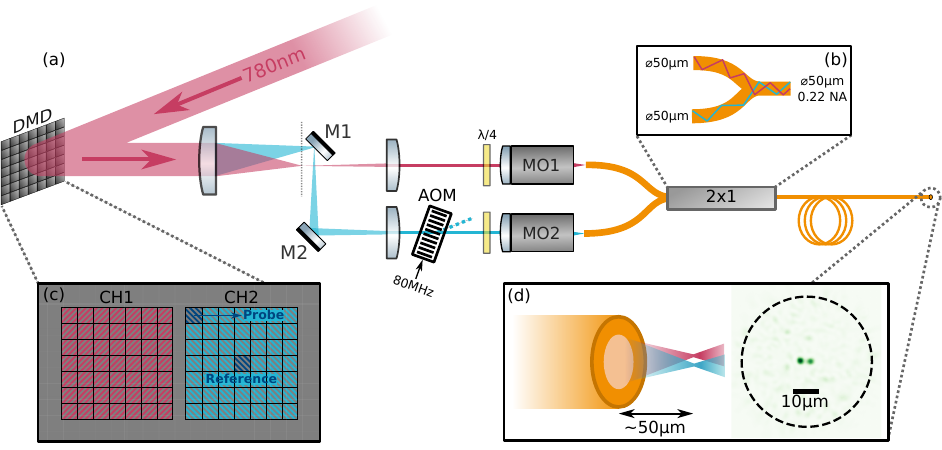}
\caption{\textbf{Wavefront shaping through a multimode-fiber coupler.} (a) A \SI{780}{\nm} laser illuminates the surface of a DMD displaying two orthogonally angled grating arrays. At the focal plane of the telescope, the two grating arrays can be spatially distinguished and directed towards their corresponding MMF inputs, allowing dual-channel control.
An AOM provides an \SI{80}{\mega\hertz} frequency offset to one of the channels to prevent interference. 
(b) Operating principle of the $2 \times 1$ fused coupler. Light from individual MMFs are merged together into an output fiber of \SI{50}{\micro m} diameter.
(c) Simplified DMD frame used when measuring projecting two channels. Each grating array is made up of $21\times21$ gratings each $16\times \SI{16}{px}$ in size. CH2 exemplifies the calibration procedure of a probe grating being sequentially scanned and compared to the central reference grating. (d) At the distal end of the MMF, a plane $\sim$\SI{50}{\micro \meter} away from the fiber facet is imaged onto a CCD camera via a $\times29$ magnification imaging system.}
\label{fig:setup}
\end{figure}

The experimental setup used in this work is shown in Fig.~\ref{fig:setup}a.
We shape light using a $2 \times 1$ fiber-optic coupler (Thorlabs TM50R5F1A) comprising of two fused multimode fibers (MMFs), which we operate in a two-input, one-output configuration (see Fig.~\ref{fig:setup}b).
The light is sourced from a distributed-feedback laser (Toptica Eagleyard) and is at wavelength $\lambda=780$\,nm.
In Sec.~\ref{sec:single-tweezer}, we use a single input to shape light into an optical tweezer (the red beam path in Fig.~\ref{fig:setup}a).
In Sec.~\ref{sec:dual-tweezer}, we use both inputs to generate two independently controllable tweezers with a frequency offset between them (i.e.\ we add the blue beam path in Fig.~\ref{fig:setup}a).

To generate the optical tweezers, we control the spatial modes launched into the MMF using a digital micromirror device (DMD, Texas Instruments DLP6500FYE), as shown in Fig.~\ref{fig:setup}c.
The incident beam onto the DMD provides approximately uniform illumination across the effective DMD area of $\SI{14.6}{\milli\meter}\times\SI{8.2}{\milli\meter}$.
After reflection from the DMD, we circularly polarize the light with quarter waveplates to suppress polarization changes as the light propagates through the MMFs~\cite{Ploschner2015}.
Then, we couple the light into the proximal facets of the MMFs using telescopes containing microscope objective lenses (Thorlabs RMS40X) with numerical apertures of 0.65.
The numerical aperture of the objectives is sufficient to fill that of the MMFs ($\mathrm{NA} = 0.22$) which have core diameters $d = \SI{50}{\micro\meter}$.
As will be discussed in Sec~\ref{sec:dual-tweezer}, when using both MMF inputs, we direct two independent beams towards independent objective lenses (labelled MO1 and MO2) which couple to the two input facets.

We form optical tweezers at a plane approximately $\SI{50}{\micro\meter}$ from the distal facet of the MMF coupler (see Fig.~\ref{fig:setup}d), set by the plane at which we characterize the transmission matrices of the fibers (see Sec.~\ref{sec:tm-measurement}). 
This plane is imaged onto a 1.6-megapixel CMOS camera (Thorlabs CS165MU/M) using an achromatic doublet lens (Comar) in a telescope that gives a magnification of approximately $29\times$. 
The choice of the distance of the tweezer plane from the distal facet is somewhat arbitrary; with our parameters, we expect diffraction-limited spots to be achievable at distances of up to approximately \SI{200}{\micro\meter} from the facet~\cite{Lee2022}.

\subsection{DMD holography}
To control the input field of the MMFs, we operate the DMD in the Lee configuration~\cite{Lee1979}.
As the micromirrors of the DMD provide only binary amplitude modulation, we divide the DMD into a grid of $16\times16$-micromirror diffraction gratings, indexed by $n$.
We encode a desired complex field $E(u,v)$ by multiplying it by a carrier $e^{i2\pi f_0\left(u\cos\theta+v\sin\theta\right)}$, corresponding to a linear phase ramp that produces distinct diffraction orders from the DMD.
The angle $\theta$ of the phase ramp determines the direction of the diffraction orders; we typically use $\theta = \pm\SI{45}{\degree}$.
Each grating, centered at $(u_n,v_n)$, is used to encode a single complex field coefficient $E_n\equiv E(u_n,v_n)=A_n e^{i\phi_n}$.

Within grating $n$, the phase $\phi_n$ is encoded through the relative phase of the carrier.
Specifically, a mirror at fine coordinate $(u,v)$ within the grating is switched on when
$\cos\!\big[\phi_n+2\pi f_0\left(u\cos\theta+v\sin\theta\right)\big]>0$
and off otherwise.
The amplitude $A_n$ is encoded by retaining a fraction $A_n$ of the grating rows in the on state, with the remaining rows switched off.
The selected rows are distributed as evenly as possible across the grating to reduce spatial bias.
After spatial filtering to retain the first diffraction order, the resulting field approximates the desired complex field, with the spatial resolution limited by the finite grating size.

\subsection{Transmission-matrix measurement} \label{sec:tm-measurement}
The propagation of light through a MMF is described by a transmission
matrix $\mathbf{T}$ \cite{Popoff2010}.
Its elements $\mathbf{T}_{mn}$ describe the complex coupling between the $n$th input mode and the $m$th output mode,
such that
\begin{equation}
  E^{\mathrm{out}}_m = \sum_n \mathbf{T}_{mn} E^{\mathrm{in}}_n .
\end{equation}
For convenience, we use the input-mode basis defined by the DMD gratings (i.e.\ each basis mode corresponds to the light diffracted from the DMD when only one grating is on) and the output-mode basis corresponding to a single point in the camera plane being illuminated.
We measure the elements of $\mathbf{T}$ using an inline holography scheme~\cite{Dubois2002,Popoff2010,Dremeau2015,Fuentes2016,Florentin2018,Zhao2021,Ivanina2026}.
This approach avoids the need for a separate reference arm.

We use an array of $21\times21$ gratings on the DMD to address the spatial modes supported by each MMF.
For our step-index MMFs, $M\approx V^2/2$ modes propagate, with
$V = 2\pi (d/2)\,\mathrm{NA}/\lambda$ \cite{Pask1975}; for a single
polarization this gives $M/2\approx491$, so our input basis spans roughly all
modes of one polarization. One grating remains static on the array and serves as a fixed-phase reference. The remaining $440$
gratings are probed sequentially (Fig.~\ref{fig:setup}c): for each probe
grating $n$ we display it alongside the reference, step its phase through
$\delta = 0,\pi/2,\pi,3\pi/2$, and record the camera image at each step. The
camera pixels are binned into $6\times6$-pixel superpixels and we retain an $80\times80$-superpixel region of interest containing the image of the fiber output.
When the reference and probe gratings are displayed together, the intensity
recorded at superpixel $m$ is $I_{mn}(\delta) = I^{\mathrm{r}}_{m} + I^{\mathrm{p}}_{mn} + 2A_{mn}\cos(\phi_{mn}+\delta)$,
where $I^{\mathrm{r}}_{m}$ and $I^{\mathrm{p}}_{mn}$ are the intensities
produced by the reference grating and by probe grating $n$ alone,
$A_{mn}=\sqrt{I^{\mathrm{r}}_{m}I^{\mathrm{p}}_{mn}}$ is the
interference-fringe amplitude, and $\phi_{mn}$ is the phase of the probe
field relative to the reference.
Thus,
\begin{equation}
  \mathbf{T}_{mn} = A_{mn}e^{i\phi_{mn}}
  = \tfrac{1}{4}\Big[\big(I_{mn}(0)-I_{mn}(\pi)\big)
  + i\big(I_{mn}(3\pi/2)-I_{mn}(\pi/2)\big)\Big],
\end{equation}
up to a factor set by the reference field at each superpixel. This factor is
independent of $n$ and only contributes a global phase when focusing onto a
single output point, so we neglect it.
We repeat the measurement for all probe modes to obtain the complete transmission matrix.
We set the column corresponding to the reference grating to zero, since this mode is not independently measured. 

\subsection{Input-field reconstruction}
The transmission matrix provides the forward mapping from the available input modes to the output field.
Experimentally, however, we require the inverse operation: given a desired output field $E^\mathrm{out}$, we seek the input coefficients $E^\mathrm{in}$ that produce the closest achievable approximation.

To approximate the desired output mode, we calculate the Moore-Penrose pseudoinverse $\mathbf{T}^{+}$ and obtain the input coefficients according to $E^\mathrm{in}=\mathbf{T}^{+}E^\mathrm{out}$. 
This is the least-squares solution, minimizing $\sum_m \left|(\mathbf{T}E^\mathrm{in})_m-E^\mathrm{out}_m\right|^2$.
The resulting input field therefore produces the closest field to the desired output that can be generated using the finite set of input modes available to the system.
Experimentally, we use only the phase component of the transmission matrix to focus the light.

\section{Generation of a single optical tweezer} \label{sec:single-tweezer}
First, we demonstrate the formation of a single optical tweezer through one input of the MMF coupler.
We use the red arm shown in Fig.~\ref{fig:setup}a and an array of $21\times21$ gratings on the DMD.
We begin by measuring the transmission matrix $\mathbf{T}$ as described above.
Then, to generate a single tweezer, we define a target output field $E^\mathrm{out}$ that is zero for all camera superpixels except at the desired tweezer location, where it is set to unity.
We calculate the input field $E^\mathrm{in} = \mathbf{T}^{+} E^\mathrm{out}$ and encode this onto the DMD with a Lee hologram.

\begin{figure}[]
\centering\includegraphics[width=\linewidth]{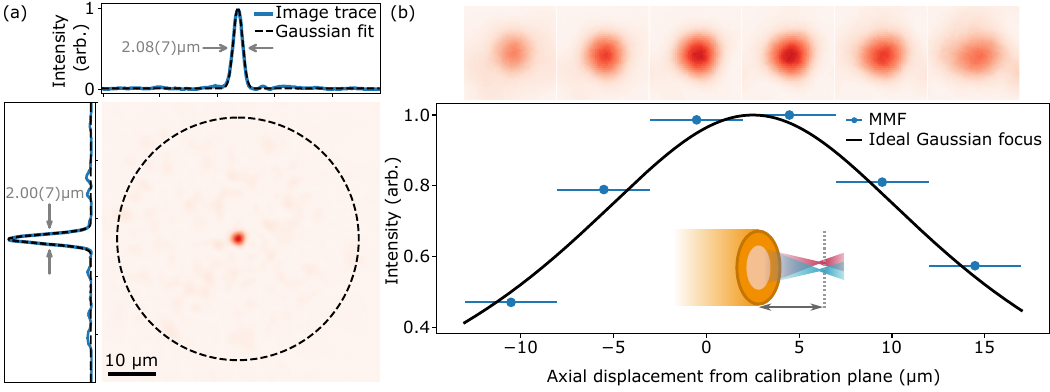}
\caption{ \textbf{Generation of a single optical tweezer through the MMF coupler.} 
(a) A camera image of an optical tweezer formed at the distal end of the MMF coupler. The dashed region shows the edge of the fiber core; we can move the tweezer around this field-of-view without significant degradation to the performance.
The top and left panels show integrations along the vertical and horizontal axes respectively.
We extract full widths at half maximum (FWHMs) of \SI{2.00(7)}{\micro\meter} and \SI{2.08(7)}{\micro\meter} along the two directions.
(b) Axial profile of the tweezer relative to the calibration plane (about \SI{50}{\micro\meter} away from the fiber) obtained by translating the fiber facet with respect to the imaging system. For comparison, the solid line shows the expected intensity profile of an ideally focused Gaussian beam through a lens of the same NA as the fiber. 
}
\label{fig:single-tweezer}
\end{figure}

Figure~\ref{fig:single-tweezer}a shows a camera image of the resulting tweezer. 
We calibrate the magnification of the imaging system using a pinhole target with known aperture sizes; the scale bar represents the scale in the plane at which the tweezer is focused.
The dashed circle indicates the extent of the coupler facet, within which the tweezer can be positioned.
We determine the tweezer waist by fitting Gaussian profiles to the intensity distributions obtained by integrating the two-dimensional image along the horizontal and vertical axes (side panels). 
The corresponding FWHMs are \SI{2.00(7)}{\micro\meter} and \SI{2.08(7)}{\micro\meter}.
These spot sizes are sufficiently small for single-atom trapping and control~\cite{Xu2026}. 
For comparison, the Abbe diffraction limit for our \SI{780}{\nano\meter} light and fiber NA of $0.22$ is $\lambda/(2\,\mathrm{NA}) \approx \SI{1.8}{\micro\meter}$~\cite{Adams2018}.

Figure~\ref{fig:single-tweezer}b shows the intensity profile of the tweezer in the direction of tweezer-light propagation. 
We take images of the tweezer by translating the fiber tip to move the tweezer through the object plane of the imaging system. The fitted intensity profile agrees well with that of an ideal diffraction-limited Gaussian focus at \SI{780}{\nm} through a lens of equivalent NA to the fiber (solid line).

\section{Generation of multiple independent optical tweezers} \label{sec:dual-tweezer}
We next demonstrate the formation of multiple optical tweezers using both inputs of the MMF coupler.
For these measurements, we use the two-input, one-output configuration of the coupler.
We control both inputs using the same DMD, with independent $21\times21$ arrays of gratings assigned to each MMF input.
The two arrays use carrier gratings with orthogonal orientations, such that the corresponding $+1$ diffraction orders are spatially separated.
Therefore, we can independently direct each order towards the corresponding objective lenses (MO1 and MO2 in Fig.~\ref{fig:setup}a) for coupling to the proximal facets of the MMF coupler.

\begin{figure}[p]
\centering\includegraphics[width=\linewidth]{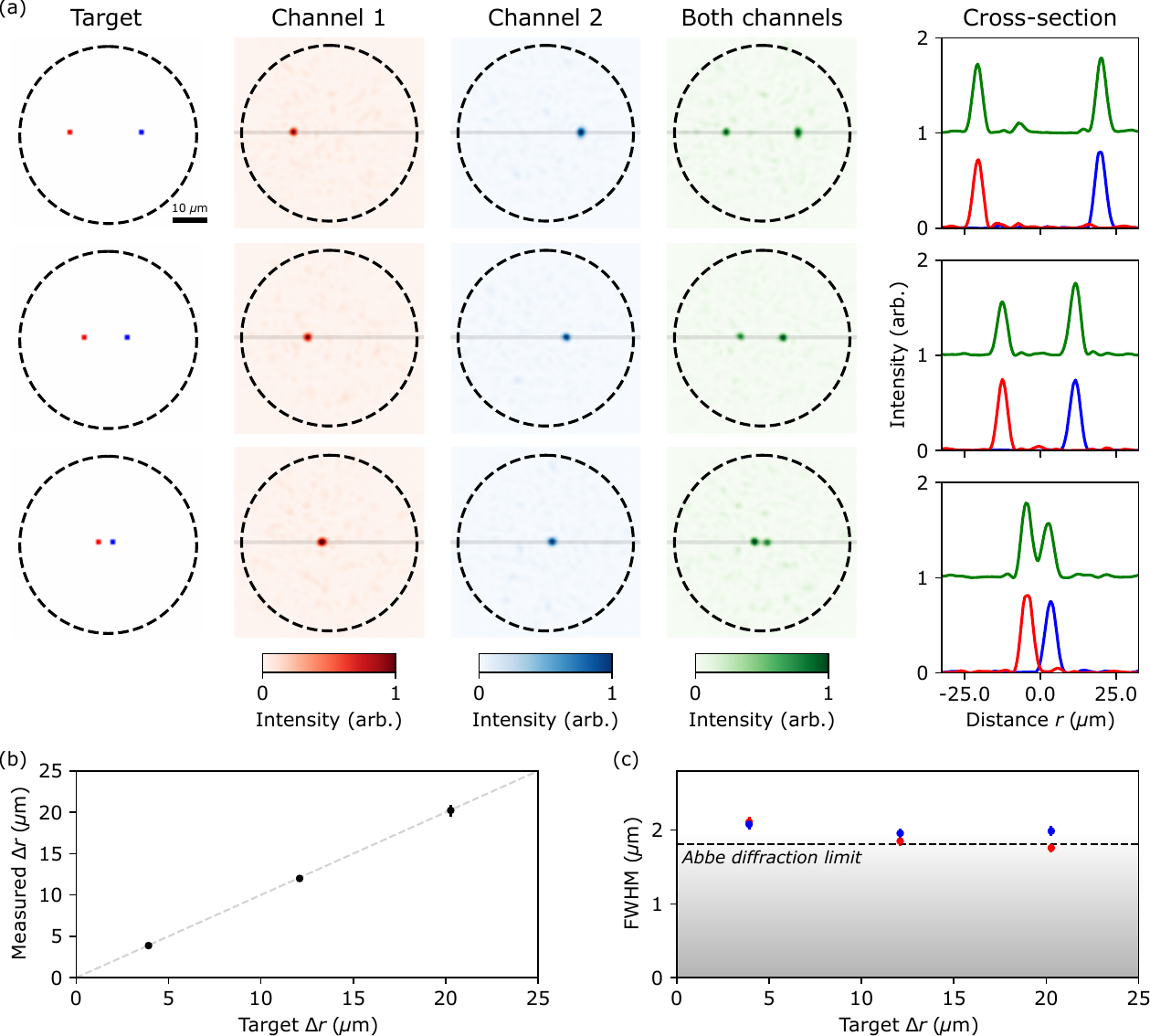}
\caption{
\textbf{Generation of multiple optical tweezers through the MMF coupler.}
(a) Examples of independent optical tweezers generated using the two input channels of the MMF coupler.
Each row corresponds to a different pair of programmed tweezer positions with the target intensities shown on the left.
In all panels, red, blue, and green correspond to the measured intensity with light coupled through channel 1, channel 2, and both channels, respectively.
The dashed circles indicate the accessible FOV of the distal facet of the MMF coupler.
The plots on the right show the corresponding intensity profiles integrated along the vertical direction in the shaded areas which we fit to determine the tweezer positions and FWHMs. We offset the dual-channel data for clarity.
(b) Control of the tweezer separation $\Delta r$. The fitted distance is in good agreement with the target distance; error bars are smaller than the size of the points.
(c) FWHMs of the tweezers as a function of the target separation. There is no significant change in performance as the traps are moved across the FOV. The black dashed line shows the Abbe diffraction limit.
}
\label{fig:dual-tweezer}
\end{figure}

We calibrate the transmission matrix of each channel independently by repeating the measurements described in Sec.~\ref{sec:tm-measurement}, using the same set of camera superpixels for both measurements.
Here, the inline holography method has a critical advantage: it is possible to easily scale the number of inputs of the MMF coupler without needing to add a large number of reference beams.
The measurements yield two transmission matrices $\mathbf{T}_c$, where $c\in\{1,2\}$ denotes the input channel.
To synthesise a target field $E^{\mathrm{out}}_c$ using channel $c$, we calculate the corresponding input coefficients $E^{\mathrm{in}}_c = \mathbf{T}_c^{+} E^{\mathrm{out}}_c$ and encode them onto the corresponding DMD grating array.
As the two grating arrays occupy disjoint sets of DMD mirrors, the two holograms can be combined into a single DMD frame without cross-talk between the channels (see Fig.~\ref{fig:setup}c).

When both channels are illuminated with coherent light of the same optical frequency, interference occurs.
To suppress the resulting time-dependent terms in the tweezer intensities, we introduce an \SI{80}{\mega\hertz} frequency offset between the two channels using an acousto-optic modulator (AOM), placed in the beam path of channel 2, as shown in Fig.~\ref{fig:setup}a.
We couple the $+1$ diffraction order from the AOM into the second input of the MMF coupler.
We compensate for the finite diffraction efficiency of the AOM: we set the intensities output from both channels to be identical by reducing the amplitude of channel 1 with its Lee hologram.
The interference term between the two optical fields oscillates at \SI{80}{\mega\hertz}, much faster than the kHz-scale trap frequencies that would be realised when trapping atoms in the optical tweezers.
Accordingly, trapped atoms would experience a time-averaged intensity given by the sum of the two independently generated tweezer intensities.

In Fig.~\ref{fig:dual-tweezer}a we demonstrate the generation of two independently controllable optical tweezers using the two input channels.
Each row corresponds to a different target configuration, as shown in the first column. 
The red and blue columns show the images when light is transmitted through only channel 1 or channel 2, respectively (the other path is blocked).
The green column shows the image when light is transmitted through both channels.
The final column shows the intensity distributions integrated within the gray regions shown in the other panels (along the vertical axis).
We observe close agreement between the measured two-channel profiles (green) and the sum of the single-channel profiles (red and blue).

To quantify the crosstalk of the two tweezers, we define the extinction ratios $X_{i,j}$ between them, in decibels, using the integrated intensities. Specifically,
\begin{equation}
    X_{1,2} = 10\log_{10}\left(\frac{I_1(r_2)}{I_1(r_1)}\right),
\end{equation}
and, similarly, $X_{2,1}$. 
Here, $I_i(r_j)$ is the intensity at the target position $r_j$ of channel $j$ when light is transmitted only through channel $i$. As $I_1(r_2)$ and $I_2(r_1)$ are at the single-count level on our 8-bit camera, we average these quantities over a \SI{1}{\micro\meter} region centered on each target position to reduce statistical fluctuations. The six extracted extinction ratios from the data in Fig.~\ref{fig:dual-tweezer} have a mean of $-21\,$dB, with a $1\sigma$ range of $-18\,$dB to $-29\,$dB. 
This corresponds to residual intensity at the percent level. 
Such intensity control is sufficient, for example, to determine which trap atoms occupy after unmerging atom pairs~\cite{Brooks2022} or to perform selective Rydberg excitation~\cite{Guttridge2023}.
Further, rather than reflecting localised crosstalk between the channels, this is consistent with the level of the random speckle background present throughout the image: integrating the background intensities \SI{10}{\micro\meter} above the tweezers (i.e.\ away from either channel's target position) gives a comparable mean extinction ratio of $-19\,$dB.

Figure~\ref{fig:dual-tweezer}b shows the measured separation between the two tweezers as a function of their target separation. We obtain the separation by fitting the integrated red and blue intensity profiles in Fig.~\ref{fig:dual-tweezer}a. The measured separation closely follows the target separation, with a fitted gradient of $0.996(3)$.

Figure~\ref{fig:dual-tweezer}c shows the FWHM of each tweezer as a function of target separation, corresponding to the tweezers being at different positions within the field of view (FOV) of the fiber. 
The dashed black line indicates the Abbe diffraction limit of approximately \SI{1.8}{\micro\meter}, with the grey shaded region indicating the regime beyond this limit. 
Across the field of view, we obtain a mean FWHM of $\SI{2.0(1)}{\micro\meter}$, where the uncertainty is the standard deviation of the measured values. 
The measured FWHM is therefore approximately $10\%$ larger than the Abbe limit. 
We attribute this to misalignment in the imaging system, which reduces the effective numerical aperture of the fiber.

\begin{figure}[]
\centering\includegraphics[width=\linewidth]{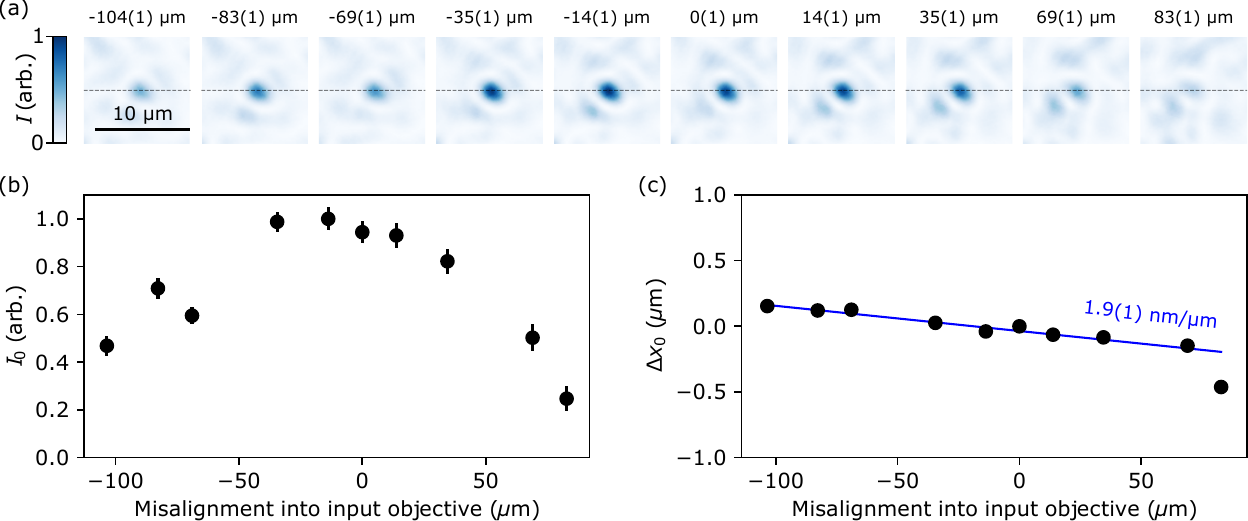}
\caption{
\textbf{Robustness of the tweezers to proximal misalignment.}
(a) Camera images of a tweezer formed through the MMF coupler for different misalignments of the beam entering the proximal objective MO2. (b,c) Tweezer amplitude  $I_0$ and positional shift $\Delta x_0$, respectively, extracted by fitting the intensity profiles along the black dashed lines in (a).
}
\label{fig:misalignment}
\end{figure}

A key advantage of the fused-coupler architecture is that the relative position of the two tweezers is intrinsically insensitive to changes in the input coupling.
A misalignment at the input does not directly translate into a displacement of the output tweezer. 
Instead, it reduces the overlap with the desired input field, transferring power into other speckle modes and consequently reducing the tweezer amplitude.
To test this, we deliberately vary the input coupling of light going into channel 2 by varying the AOM frequency. 
The change in angle of the AOM diffraction order results in the light going into MO2 being misaligned.
Example images for different misalignments are shown in Fig.~\ref{fig:misalignment}a.
As a function of the input misalignment, we measure the resulting tweezer amplitude $I_0$ (Fig.~\ref{fig:misalignment}b) and positional shift $\Delta x_0$ (Fig.~\ref{fig:misalignment}c). 
Importantly, the measured position of the tweezer remains approximately constant (with a gradient of magnitude \SI{1.9(1)}{\nano\meter\per\micro\meter}) whilst the amplitude decreases with increasing misalignment. 

\section{Discussion and Conclusion} \label{sec:discussion}
We have shown that a $2 \times 1$ MMF fused coupler can be used to generate independently controlled optical tweezers at the distal end of a MMF.
Using a single input channel, we characterised the transmission matrix of the MMF coupler and used it to form an optical tweezer.
Extending this to both channels of the coupler, we generated two simultaneous, independently addressable tweezers, and showed that their position is robust to misalignment of the input coupling.
Using the inline holographic approach to transmission-matrix measurement scales straightforwardly to additional input channels, since each new channel requires only an additional grating array on the DMD rather than a new reference beam.

For convenience, the demonstrations in this work have been performed with low optical power, approximately \SI{50}{\micro\watt} per tweezer. 
This is within the regime required for trapping atoms.
For example, to form a trap of depth \SI{100}{\micro\kelvin} for potassium atoms (i.e.\ a depth sufficient to load the atoms from a magneto-optical trap), one requires approximately $\SI{250}{\micro\watt}$ per tweezer.
Here, we have assumed that the tweezer is formed from light at wavelength \SI{780}{\nano\meter} (the wavelength used throughout this work) with a FWHM of $\SI{2.1}{\micro\meter}$, and have taken the polarizability of the atom to be $8720\times4\pi\varepsilon_0 a_0^3$~\cite{Kiruga2026}.
The power requirements could be reduced further by decreasing the size of the tweezers, either by choosing a species trappable at shorter wavelength (e.g.\ ytterbium at \SI{532}{\nano\meter} \cite{saskin2019narrow}) or by using a fiber with a higher NA~\cite{Vitrant2021,Garcia2022}.

Scaling this approach to larger arrays will require higher total optical power than is used in this work. 
There are many readily available laser sources with output powers on the order of several watts that would be suitable for large arrays.
A more intriguing prospect is to amplify the light within the multimode fiber itself: Florentin \textit{et al.}~\cite{Florentin2017} demonstrated MMF-amplifier-based beam shaping achieving approximately \SI{1}{\watt} of shaped light at \SI{1064}{\nano\meter}. This would allow beam shaping to be performed at low power before the light is amplified to form the atomic potentials, which may be beneficial in deployment environments.

The MMF coupler is well suited to the amplitude-control requirements of atom-based quantum technologies, since isolating the channels used to form each tweezer allows their intensities to be controlled independently. 
For example, atoms can be selectively excited to Rydberg states by switching off the corresponding trap to remove the trap-induced AC Stark shift. 
This technique is widely used in Rydberg-atom experiments and requires switching on the $\si{\micro\second}$ timescale~\cite{Labuhn2016}. 
Because each tweezer is formed from a distinct input channel, adding an AOM to each channel would allow either trap to be switched on this timescale, faster than the refresh rates typically achievable with DMDs ($\sim\si{\kilo\hertz}$). 
In addition to amplitude control, DMD-controlled potentials allow atoms to be moved, which is important for flexible atomic connectivity~\cite{Bluvstein2022} and for rearranging atoms to form defect-free arrays~\cite{Miroshnychenko2006,Lee2016,Barredo2016,Endres2016}. 
Such movement has previously been demonstrated using a focusing lens~\cite{Stuart2018}, and we expect the same approach to be applicable to the MMF coupler.
The compact size and flexibility of the MMF coupler may also make it suitable for integration into cryogenic environments, where optical access is limited. 

Additionally, the robustness of the tweezers to relative misalignment is highly advantageous for studies requiring close (sub-waist) and stable separations of atoms and molecules. For example, studies of weak interactions between atoms and molecules require particles to be prepared at distances on the 100-nm scale, where shot-to-shot noise in the trap separation causes dephasing~\cite{Guttridge2023}. Positional stability on this scale may also be required for future studies of interactions between magnetic atoms~\cite{du2024atomic}.

More broadly, to the best of our knowledge, this work represents the first demonstration of beam shaping through a multimode-fiber coupler, and the two-input, one-output configuration we have used could be beneficial in a range of settings beyond the atomic-physics context considered here. For example, microparticles could be trapped~\cite{Bianchi2012} in a potential formed through one input channel while being imaged through the other. Inverting the device to a one-input, two-output configuration could allow a diagnostic endoscope~\cite{Cao2023,Amitonova2024} to image two regions simultaneously from a single fiber. The same principle extends naturally to optical communications: multiple coherent, independently steerable channels sharing a common fiber output are well suited to space-division multiplexing~\cite{Richardson2013}, offering a route to increasing data throughput without adding fibers.

These results establish MMF couplers as a compact and scalable route to independently controllable optical fields, extending fiber-based approaches beyond the single-channel operation demonstrated previously. Combined with the power requirements, amplitude control, and coupling insensitivity discussed above, this architecture offers a practical platform for reconfigurable, addressable tweezer arrays in low-SWaP-C environments. The natural next step is to trap and image single atoms in tweezers generated through an MMF coupler, building directly on the results presented here towards deployable atomic quantum technologies.

\section{Back matter}

\begin{backmatter}
\bmsection{Funding}
[To be generated by Prism.]

\bmsection{Acknowledgment}
We thank Ifan G. Hughes and Danielle Pizzey for the use of laboratory space, and IGH, DP, and Tobias Franzen for helpful discussions.
DKR acknowledges a Durham University \textit{Developing Talent Award} which supported the cost of equipment used in this project.

\bmsection{Disclosures}
The authors declare no conflicts of interest.

\bmsection{Data availability} Data underlying the results presented in this paper are available in Ref. [to be added].

\end{backmatter}


\end{document}